\documentclass[12pt]{iopart}
\usepackage{graphicx}
\begin{document}

\title{Solar models and solar neutrino oscillations}

\author{John N. Bahcall and Carlos Pe\~na-Garay}

\address{Institute for Advanced Study, School of Natural Sciences,
  Princeton, NJ 08540, USA}

\begin{abstract}
We provide a summary  of the current knowledge, theoretical and
experimental, of solar neutrino fluxes and of the masses and
mixing angles that characterize solar neutrino oscillations. We
also summarize the principal reasons for doing new solar neutrino
experiments and what we think may be learned from the future
measurements.
\end{abstract}

\maketitle

\section{Introduction}
\label{sec:introduction}

 We record in this paper a snapshot (taken on March 1, 2004) of
where we stand with solar neutrino theoretical research.  We do
not attempt to review the many papers written on this subject. For
details of the extensive literature, the reader is referred to
earlier, more comprehensive
studies~\cite{pontecorvo,msw,book,bp00,cabibbo02,fiorentini02,conchayossi,smirnov,roadmap,bargerreview,
bilenky03,kayser03,murayama,analysispostkamland,valle03,chitre01,sackmann03,turckchieze,haxton04}.

The related subject of solar neutrino experiments will be reviewed
in this volume by A. McDonald~\cite{mcdonald}. We therefore do not
discuss the experimental aspects of solar neutrino research in
this article, although we do emphasize the relation between
theoretical ideas and predictions and solar neutrino measurements.

We begin in Section~\ref{sec:models} by summarizing our current
theoretical knowledge of the solar neutrino fluxes.  We then
summarize in Section~\ref{sec:parameters} the numerical results
regarding solar neutrino parameters and neutrino fluxes that have
been inferred from solar neutrino and reactor experiments.
Neutrinos are the first cosmological dark matter to be discovered.
We describe in Section~\ref{sec:darkmatter} what solar and
atmospheric neutrino experiments have taught us about the
cosmological mass density in neutrinos. Finally, in
Section~\ref{sec:newexperiments} we discuss the reasons for doing
future solar neutrino experiments and the scientific results that
may be obtained from the proposed new experiments.

\section{Solar Model Fluxes}
\label{sec:models}

We base the discussion in this section on the results reported in
the recent paper~\cite{bp04}. Full numerical details of the solar
models, BP04 and BP00 that are discussed below are presented,
together with earlier solar models in this series, at the Web
site: http://www.sns.ias.edu/$\sim$jnb .

\begin{figure}
\begin{center}
\includegraphics [width=4.5in,angle=270]{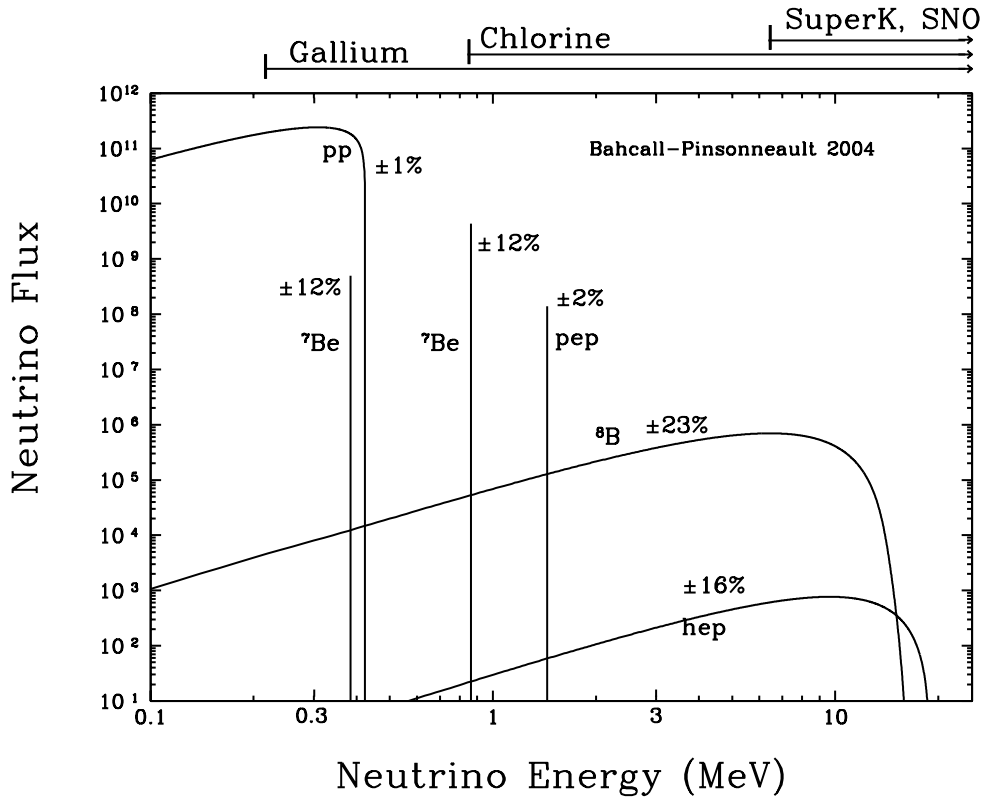}
\end{center}
\caption{The predicted solar neutrino energy spectrum. The figure
shows the energy spectrum of solar neutrinos predicted by the BP04
solar model~\cite{bp04}.  For continuum sources, the neutrino
fluxes are given in number per ${\rm cm^{-2} sec^{-1} MeV^{-1}}$
at the Earth's surface. For line sources, the units are number per
${\rm cm^{-2} sec^{-1}}$. The total theoretical uncertainties
taken from column 2 of Table~\ref{tab:neutrinofluxes} are shown
for each source. In order not to complicate the figure, we have
omitted the difficult-to-detect CNO neutrino fluxes (see
Table~\ref{tab:neutrinofluxes}). \label{fig:bp04} }
\end{figure}

\subsection{Fluxes from different solar models}
\label{subsec:fluxes}

\begin{table}[ht]
\caption{Predicted solar neutrino fluxes from solar models. The
table presents the predicted fluxes, in units of $10^{10}(pp)$,
$10^{9}({\rm \, ^7Be})$, $10^{8}(pep, {\rm ^{13}N, ^{15}O})$,
$10^{6} ({\rm \, ^8B, ^{17}F})$, and $10^{3}(hep)$ ${\rm
cm^{-2}s^{-1}}$. Columns 2-4 show BP04, BP04+, and the previous
best model BP00~\cite{bp00}. Columns 5-7 present the calculated
fluxes for solar models that differ from  BP00  by an improvement
in one set of input data: nuclear fusion cross sections (column
5), equation of state for the solar interior (column 6), and
surface chemical composition for the Sun (column 7). Column~8 uses
the same input data as for BP04 except for a recent report of the
$^{14}$N + p fusion cross section. References to the improved
input data are given in the text.  The last two rows ignore
neutrino oscillations and present for the chlorine and gallium
solar neutrino experiments the capture rates in SNU (1 SNU equals
$10^{-36}{\rm ~events ~per~target~atom~per~sec}$). Due to
oscillations, the measured rates are smaller: $2.6 \pm 0.2$ and
$69 \pm 4$, respectively. The neutrino absorption cross sections
and their uncertainties are given in Ref.~\cite{nucrosssections}.
 \protect\label{tab:neutrinofluxes}}
\lineup
\begin{indented}
\item[]\begin{tabular}{@{}lccccccc}
\br
Source&\multicolumn{1}{c}{BP04}&{BP04+}&BP00&Nucl&EOS&Comp&$^{14}$N\\
\mr
$pp$&5.94$(1 \pm 0.01)$&5.99 &5.95&5.94&5.95&6.00&5.98\\
$pep$&1.40$(1 \pm 0.02) $&1.42&1.40&1.40&1.40&1.42&1.42\\
$hep$&$7.88 (1 \pm 0.16)$&8.04&9.24&7.88&9.23&9.44&7.93\\
${\rm ^7Be}$&$4.86 (1 \pm 0.12)$&4.65&4.77&4.84&4.79&4.56&4.86\\
${\rm ^8B}$&5.82$(1 \pm 0.23)$&5.28&5.05&5.79&5.08&4.62&5.77\\
${\rm ^{13}N}$&$5.71(1~~^{+0.37}_{-0.35}) $&4.06&5.48&5.69&5.51&3.88&3.23\\
${\rm ^{15}O}$&$5.03(1~~^{+0.43}_{-0.39}) $&3.54 &4.80&5.01&4.82&3.36&2.54\\
${\rm ^{17}F}$&$5.91(1~~^{+0.44}_{-0.44}) $&3.97&5.63&5.88&5.66&3.77&5.85\\
\noalign{\smallskip} \hline \noalign{\smallskip}
Cl&$8.5^{+1.8}_{-1.8}$&7.7&7.6&8.5&7.6&6.9&8.2\\
Ga&131$^{+12}_{-10}$&126&128&130&129&123&127\\
\br
\end{tabular}
\end{indented}
\end{table}

Table~\ref{tab:neutrinofluxes}, taken from Ref.~\cite{bp04}, gives
the calculated solar neutrino fluxes for a series of solar models
calculated with different plausible assumptions about the input
parameters. The range of fluxes shown for these models illustrates
the systematic uncertainties in calculating solar neutrino fluxes.
The second (third) column, labelled BP04 (BP04+), of
Table~\ref{tab:neutrinofluxes} presents the current best solar
model calculations for the neutrino fluxes. The uncertainties are
given in column~2.

Figure~\ref{fig:bp04} presents the neutrino energy spectrum
predicted by the BP04 solar model for the most important solar
neutrino sources.

The model BP04+ was calculated with the use of new input data for
the equation of state, nuclear physics, and solar composition. The
model BP04, the currently preferred model, is the same as BP04+
except that BP04 does not include the most recent analyses of the
solar surface composition~\cite{newcomp}, which conflict with
helioseismological measurements. We prefer the model BP04 over the
model BP04+ because the lower heavy element abundance used in
calculating BP04+ causes the calculated depth of the solar
convective zone to conflict with helioseismological measurements.

The error estimates, which are the same for the three models
labeled BP04, BP04+, and $^{14}$N in
Table~\ref{tab:neutrinofluxes}) include the recent composition
analyses.

Column four of Table~\ref{tab:neutrinofluxes} presents the fluxes
calculated using the preferred solar model, BP00~\cite{bp00}, that
was posted on the archives in October 2000. The BP04 best-estimate
neutrino fluxes and their uncertainties have not changed markedly
from their BP00 values despite refinements in input parameters.
The only exception is the CNO flux uncertainties which have almost
doubled   due to the larger systematic uncertainty in the surface
chemical composition estimated in this paper.

We describe   improvements in the input data relative to BP00.
Quantities that are not discussed here are the same as for BP00.
Each class of improvement is represented by a separate column,
columns 5-7,  in Table~\ref{tab:neutrinofluxes}. The magnitude of
the changes between the fluxes listed in the different columns of
Table~\ref{tab:neutrinofluxes} are one measure of the sensitivity
of the calculated fluxes to the input data.

 Column~5 contains the fluxes
computed for a solar model that is identical to BP00 except that
improved values for direct measurements of the
$^7$Be(p,$\gamma$)$^8$B cross section~\cite{b8,pphep}, and the
calculated p-p and hep cross sections~\cite{pphep}. The reactions
that produce the $^8$B and hep neutrinos are rare; changes in
their production cross sections only affect, respectively, the
$^8$B and hep fluxes. The 15\% increase in the calculated $^8$B
neutrino flux, which is primarily due to a more accurate cross
section for $^7$Be(p,$\gamma$)$^8$B, is the only significant
change in the best-estimate fluxes.

The fluxes in Column~6 were calculated using a refined equation of
state, which includes relativistic corrections and a more accurate
treatment of molecules~\cite{eos}. The equation of state
improvements between 1996 and 2001, while significant in some
regions of parameter space, change all the solar neutrino fluxes
by less than 1\%. Solar neutrino calculations are insensitive to
the present level of uncertainties in the equation of state.

The most important changes in the astronomical data since BP00
result from new analyses of the surface chemical composition of
the Sun. The input chemical composition affects the radiative
opacity and hence the physical characteristics of the solar model,
and to a lesser extent the nuclear reaction rates. New values for
C,N,O,Ne, and Ar have been derived~\cite{newcomp} using
three-dimensional rather than one-dimensional atmospheric models,
including hydrodynamical effects, and paying particular attention
to uncertainties in atomic data and observational spectra. The new
abundance estimates, together with the previous best-estimates for
other solar surface abundances~\cite{oldcomp}, imply a ratio of
heavy elements to hydrogen by mass of $Z/X = 0.0176$, much less
than the previous value of $Z/X = 0.0229$~\cite{oldcomp}. Column~7
gives the fluxes calculated for this new composition mixture. The
largest change in the neutrino fluxes for the  p-p chain is the
9\% decrease   in the predicted $^8$B neutrino flux. The N and O
fluxes are decreased by much more, $\sim 35 \%$, because they
reflect directly the inferred C and O abundances.

The CNO nuclear reaction rates are less well determined than the
rates for the more important (in the Sun) p-p
reactions~\cite{adelberger}. The rate for
$^{14}$N(p,$\gamma$)$^{15}$O is poorly known, but important for
calculating CNO neutrino fluxes.  Extrapolating to the low
energies relevant for solar fusion introduces a large uncertainty.
Column 8 gives the neutrino fluxes calculated with input data
identical to BP04  except for the  cross section factor $S_0({\rm
^{14}N + p}) = 1.77 \pm 0.2\, {\rm keV~b}$ that is about half the
current best-estimate;  this value assumes a particular R-matrix
fit to the experimental data~\cite{new14n}. The p-p cycle fluxes
are changed by only $\sim 1$\%, but the $^{13}$N and $^{15}$O
neutrino fluxes are reduced by $ 40\%-50$\% relative to the BP04
predictions. CNO nuclear reactions contribute 1.6\% of the solar
luminosity in the BP04 model and only 0.8\% in the model with a
reduced $S_0({\rm ^{14}N + p})$.

\subsection{Flux uncertainties}
\label{subsec:fluxuncertainties}

Table~\ref{tab:uncertainties}, also taken from Ref.~\cite{bp04},
shows the individual contributions to  the flux uncertainties.
These uncertainties are useful in deciding how accurately we need
to determine  a given input parameter should be determined.
Moreover, the theoretical flux uncertainties continue to play a
significant role in some determinations of neutrino parameters
from solar neutrino experiments (see, e.g.,
Ref.~\cite{postkamland}).

\begin{table}[ht]
\caption[]{Principal sources of uncertainties in calculating solar
neutrino fluxes.
 Columns 2-5 present the fractional uncertainties in the neutrino fluxes from laboratory
 measurements of, respectively, the $^3$He-$^3$He, $^3$He-$^4$He, p-$^7$Be, and p-$^{14}$N
 nuclear fusion reactions. The last four columns, 6-9, give, respectively, the fractional uncertainties
 due to  the calculated radiative opacity, the calculated rate of element diffusion,
 the measured solar luminosity, and the measured heavy element to hydrogen ratio.\protect\label{tab:uncertainties}}
\lineup
\begin{indented}
\item[]\begin{tabular}{lccccccccc} \br
Source&\multicolumn{1}{c}{3-3}&3-4&1-7&1-14&Opac&Diff&$L\odot$&Z/X\\
\mr
$pp$&0.002&0.005&0.000&0.002&0.003&0.003&0.003&0.010\\
$pep$&0.003&0.007&0.000&0.002&0.005&0.004&0.003&0.020\\
$hep$&0.024&0.007&0.000&0.001&0.011&0.007&0.000&0.026\\
${\rm ^7Be}$&0.023&0.080&0.000&0.000&0.028&0.018&0.014&0.080\\
${\rm ^8B}$&0.021&0.075&0.038&0.001&0.052&0.040&0.028&0.200\\
${\rm ^{13}N}$&0.001&0.004&0.000&0.118&0.033&0.051&0.021&0.332\\
${\rm ^{15}O}$&0.001&0.004&0.000&0.143&0.041&0.055&0.024&0.375\\
${\rm ^{17}F}$&0.001&0.004&0.000&0.001&0.043&0.057&0.026&0.391\\
\br
\end{tabular}
\end{indented}
\end{table}

Columns~2-5 present the fractional uncertainties from the nuclear
reactions whose measurement errors are most important for
calculating neutrino fluxes. Unless stated otherwise, the
uncertainties in the nuclear fusion cross sections are taken from
Ref.~\cite{adelberger}.

The measured rate of the $^3$He-$^3$He reaction, which
 changed by a factor of 4 after the first solar
model calculation of the solar neutrino flux~\cite{series}, and
the measured rate of the $^7$Be + p reaction, which for most of
this series has been the dominant uncertainty in predicting the
$^8$B neutrino flux, are by now very well determined. If the
current published systematic uncertainties for the $^3$He-$^3$He
and $^7$Be + p reactions are correct,then the uncertainties in
these reactions no longer contribute in a crucial way to the
calculated theoretical uncertainties (see column~2 and column~4 of
Table~\ref{tab:uncertainties}). This felicitous situation is the
result of an enormous effort extending over four decades, and
represents a great collective triumph, for the nuclear physics
community.

 At the present time, the most important nuclear
physics uncertainty in calculating solar neutrino fluxes is  the
rate of the $^3$He-$^4$He reaction (see column~3 of
Table~\ref{tab:uncertainties}).  The systematic uncertainty in the
the rate of $^3$He($^4$He, $\gamma$)$^7$Be reaction(see
Ref.~\cite{adelberger})  causes an 8\% uncertainty in the
prediction of both the $^7$Be and the $^8$B solar neutrino fluxes.
It is scandalous that there has not been any progress in the past
15 years in measuring this rate more accurately.

For $^{14}$N(p,$\gamma$)$^{15}$O, we have continued to use in
Table~\ref{tab:uncertainties} the uncertainty given in
Ref.~\cite{adelberger}, although the recent reevaluation in
Ref.~\cite{new14n} suggests that the uncertainty could be somewhat
larger (see column~7 of Table~\ref{tab:neutrinofluxes}).

The uncertainties due to the calculated radiative opacity and
element diffusion, as well as the measured solar luminosity
(columns 6-8 of Table~\ref{tab:uncertainties}), are all moderate,
non-negligible but not dominant. For the $^8$B and CNO neutrino
fluxes, the uncertainties that are due to the radiative opacity,
diffusion coefficient, and solar luminosity are all in the range
2\% to 6\%.

The surface composition of the Sun is the most problematic and
important source of uncertainties. Systematic errors dominate: the
effects of line blending, departures from local thermodynamic
equilibrium, and details of the model of the solar atmosphere. In
the absence of detailed information to contrary, it is  assumed
that the uncertainty in all important element abundances is
approximately the same. The $3\sigma$ range of $Z/X$ is defined as
the spread over all modern determinations (see
Refs.~\cite{book,bp00,series}), which  implies that at present
$\Delta (Z/X)/(Z/X) = 0.15 ~(1\sigma)$, 2.5 times larger than the
uncertainty adopted in in discussing the predictions of the model
BP00~\cite{bp00}. The most recent uncertainty quoted for oxygen,
the most abundant heavy element in the Sun, is similar: 12\%
\cite{newcomp}.

Heavier elements like Fe affect the radiative opacity and hence
the neutrino fluxes more strongly than the relatively light
elements~\cite{bp00}.  This is the reason why the difference
between the fluxes calculated with BP04 and BP04+ (or between BP00
and Comp, see Table~\ref{tab:neutrinofluxes}) is less than would
be expected for the 26\% decrease in $ Z/X$. The abundances that
have changed significantly since BP00 (C, N, O, Ne, Ar) are all
for lighter species for which meteoritic data are not available.

 The dominant uncertainty listed in Table~\ref{tab:uncertainties} for
the $^8$B and CNO neutrinos is  the chemical composition,
represented by $Z/X$ (see column~9). The uncertainty ranges from
20\% for the $^8$B neutrino flux to $\sim 35$\% for the CNO
neutrino fluxes. Since the publication of BP00, the best published
estimate for Z/X decreased by $4.3\sigma$(BP00 uncertainty) and
the estimated uncertainty due to $Z/X$ increased for $^8$B
($^{15}$O) neutrinos by a factor of 2.5 (2.8). Over the past three
decades, the changes have almost always been toward a smaller
$Z/X$. The monotonicity  is surprising since different sources of
improvements have caused successive changes. Nevertheless, since
the changes are monotonic, the  uncertainty estimated from the
historical record is large.

\section{Experimentally Determined Solar Neutrino Parameters}
\label{sec:parameters}

\subsection{Solar Neutrino Oscillations}
\label{subsec:oscillations}

Solar neutrino experiments have demonstrated solar neutrinos undergo flavor conversion.
Recently, the mechanism of conversion has been identified as neutrino oscillations, i.e.,
flavor conversion induced by neutrino masses and mixing angles. A triumph of several decades
of research in solar neutrinos has been the confirmation of the predicted neutrino oscillation
deficit observed in the Japanese reactor (anti)neutrino detector KamLAND~\cite{kamlandfirstpaper}.

The Standard Model of particle physics has to be extended to include neutrino masses and mixing
angles. Oscillation experiments are sensitive to mixing angles, defined by the non trivial relation
between flavor and mass neutrino fields. Oscillation experiments are not sensitive to absolute masses
but to the differences of squared masses, i.e., global phases are not observable, relative phases are
observable. A detailed discussion of the space of oscillation parameters can be found in \cite{3par}.

Solar neutrino oscillations are characterized by just one
function, the survival probability of electron neutrinos :
neutrino production, evolution and detection are equally sensitive
to muon and tau neutrinos. The survival probability of electron
neutrinos, $P_{ee}$, can be related to the survival probability,
$P_{\rm ee}^{2\nu}$, for effective two neutrino oscillations by
the equation~\cite{kuo,3nu}

\begin{equation}
P_{ee}~=~\cos^4\theta_{13} P^{2\nu}_{ee}(\Delta m^2, \theta_{12} ; \cos^2\theta_{13} n_e)
+ \sin^4\theta_{13}.
 \label{eq:three}
\end{equation}
Here $\Delta m^2$ and $\theta_{1i}$ are, respectively, the
difference in the squares of the masses of the two neutrinos and
the vacuum mixing angles. The  effective two-neutrino problem is
solved with a rescaled electron density, $\cos^2\theta_{13} n_e$.
The effect of $\Delta M^2$, the mass difference squared
characteristic of atmospheric neutrinos, averages out  in
Equation~\ref{eq:three} for the energies and distances
characteristic of solar neutrino propagation. The results from the
CHOOZ reactor experiment~\cite{chooz,paloverde} place a strong
upper bound on $\sin^2 2\theta_{13}$, implying that $\theta_{13}$
is close to $0$ or close to $\pi/2$. Atmospheric and solar data
select the first option ($\sin^2\theta_{13} < 0.052$ at
$3\sigma$~\cite{3nuupdate}). Thus the main effect of a small
allowed $\theta_{13}$ on the survival probability is the
introduction of the factor $\cos^4\theta_{13}$ in
Equation~\ref{eq:three}.

The effective Hamiltonian for two-neutrino propagation in matter can
be written conveniently in the familiar
form~\cite{msw,book,conchayossi,bethe,Mikheev:ik,messiah,kuo}

\begin{equation}
H ~=~ \left ( \begin{array}{cc} \frac{\Delta m^2 cos 2 \theta_{12}}{4 E}- \frac{\sqrt{2}G_{\rm F} \cos^2\theta_{13} n_{\rm e}}{2}&
\frac{\Delta m^2 sin2 \theta_{12}}{2 E}\\
 \frac{\Delta m^2 sin2 \theta_{12}}{2 E} &
 -\frac{\Delta m^2 cos 2 \theta_{12}}{4 E}+ \frac{\sqrt{2}G_{\rm F} \cos^2\theta_{13} n_{\rm e}}{2}\end{array}\right) \, .
 \label{eq:hamiltonian}
\end{equation}

Here $E$ is the energy of the neutrino, $G_{\rm F}$ is the Fermi coupling constant. The  relative
importance of the MSW matter term and the kinematic vacuum oscillation term in the
Hamiltonian can be parameterized by the quantity, $\beta$, which represents the ratio of matter to vacuum
effects. From Equation~\ref{eq:hamiltonian} we see that the appropriate ratio is
\begin{equation}
\beta~=~ \frac{2 \sqrt2 G_F \cos^2\theta_{13} n_e E_\nu}{\Delta
m^2}\, . \label{eq:defbeta}
\end{equation}
The quantity $\beta$ is the ratio between the oscillation length in matter and the oscillation
 length in vacuum. In convenient units,
$\beta$ can be written as
\begin{equation}
\beta~=~ 0.22 \, \cos^2\theta_{13} \, \left[\frac{E_\nu}{1~{\rm
MeV}}\right]\, \left[ \frac{\mu_e\rho}{100~{\rm g~cm}^{-3}}\right]
\, \left[ \frac{7 \times 10^{-5} eV^2}{\Delta m^2}\right]\, ,
\label{eq:betaconvenient}
\end{equation}
where $\mu_e$ is the electron mean molecular weight ($\mu_e
\approx 0.5(1 + X)$, where X is the mass fraction of hydrogen) and
$\rho$ is the total density. For the electron density at the
center of the standard solar model, $\beta = 0.22$ for $ E =
1$MeV, $\theta_{13} =0$, and $\Delta m^2 =  7\times 10^{-5} {\rm
eV^2}$.

\subsection{The Vacuum-Matter transition}
\label{subsec:transition}
 For the large mixing angle (LMA) region ($\Delta m^2 > 10^{-5} {\rm eV^2}$),
the daytime survival probability can be written to a  good approximation in the
following simple form~\cite{msw,book,conchayossi,3nu,bethe,Mikheev:ik,messiah}
\begin{equation}
P_{\rm ee}~=~\cos^4\theta_{13} (\frac{1}{2} ~+~  \frac{1}{2} \cos2\theta^M_{12} \cos2\theta_{12})\, , \label{eq:plmaday}
\end{equation}
where the mixing angle in matter is
\begin{equation}
\cos2\theta^M_{12} = \frac{\cos2\theta_{12}-\beta}{\sqrt{(\cos2\theta_{12}-\beta)^2+\sin^22\theta_{12}}}\, .
\label{eq:defthetaM}
\end{equation}

In Equation~\ref{eq:defthetaM}, $\beta$ is calculated at the
location where the neutrino is produced. The evolution is
adiabatic, i.e., the parameters in the Hamiltonian vary slowly
enough to allow the created neutrino to follow the changing
Hamiltonian eigenstate. Thus, the survival probability depends on
the initial and final density but not on details of the density
profile.

\begin{figure}
\begin{center}
\includegraphics [width=3.5in]{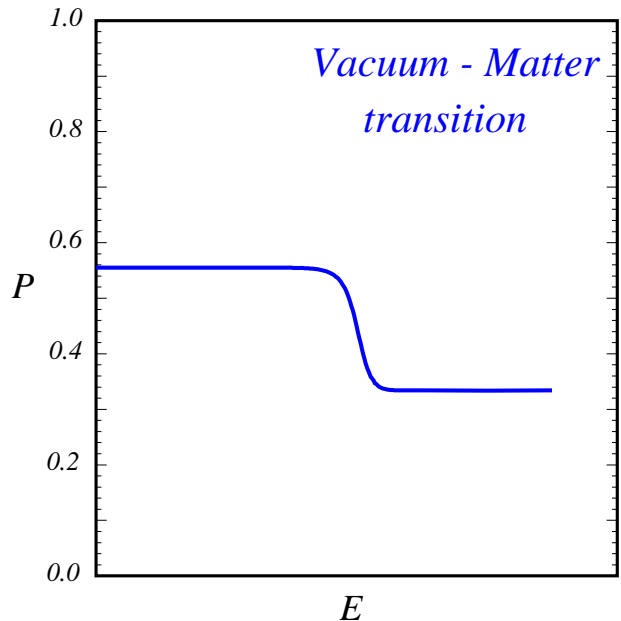}
\end{center}
\caption{\label{fig:survival} The figure shows the electron neutrino survival probability,
$P_{\rm ee}$, as a function of neutrino energy for the (daytime) LMA oscillation solution.
For small values of the parameter $\beta$ defined in Equation~\ref{eq:defbeta} and
Equation~\ref{eq:betaconvenient}, the kinematic (vacuum) oscillation effects are dominant.
For values of $\beta$ greater than unity, the MSW (matter) oscillations are most important.
For solar conditions, the transition between vacuum and matter oscillations occurs
somewhere in the region of 2 MeV.}
\end{figure}

Figure~\ref{fig:survival} illustrates the energy dependence of the LMA survival probability, $P_{\rm
ee}$. If $\beta < \cos 2\theta_{12} \sim 0.4$ (for solar neutrino oscillations), the  survival probability
corresponds to vacuum averaged oscillations,
\begin{equation}
P_{\rm ee}~=~\cos^4\theta_{13} ~(1 - \frac{1}{2} \sin^22\theta_{12}) ~\, (\beta < \cos 2\theta_{12}, ~{\rm vacuum})
.\label{eq:peevacuum}
\end{equation}
If $\beta > 1$, the  survival probability corresponds to matter dominated oscillations,
\begin{equation}
P_{\rm ee}~=~\cos^4\theta_{13} ~ \sin^2\theta_{12}~\, (\beta > 1, ~{\rm MSW}).
\label{eq:peeadiab}
\end{equation}

The survival probability is approximately constant in either of
the two limiting regimes, $\beta < \cos 2\theta_{12}$ and $\beta >
1$. The LMA solution exhibits strong energy dependence only in the
transition region between the limiting regimes. The quantity
$\beta$ is defined by Equation~(\ref{eq:defbeta}) and
Equation~(\ref{eq:betaconvenient}).

At what neutrino energy does the transition take place between vacuum oscillations and matter
oscillations? The answer to this question depends upon which neutrino source one discusses, since the
fraction of the neutrino flux that is produced at a given radius (i.e., density and $\mu_e$) differs from
one neutrino source to another. The $^8$B neutrinos are produced at much smaller radii (higher densities)
than the $p-p$ neutrinos; the $^7$Be production profile is intermediate between the $^8$Be and $p-p$
neutrinos.  According to the BP00 solar model, the critical energy at which $\beta = \cos 2\theta_{12}$
is, for $\tan^2 \theta_{12} = 0.41$,
\begin{equation}
E({\rm crit})~\simeq 1.8 {\rm ~ MeV\,(^8B)}; ~~\simeq 2.2 {\rm ~ MeV \,(^7Be)}; ~~\simeq 3.3 {\rm ~ MeV
\,}(p-p).
 \label{eq:critical}
\end{equation}

The actual energies for $p-p$ and $^7$Be neutrinos are below the critical energy where they are produced.
To a very good approximation, $^8$B neutrinos are always in the MSW regime
(Equation~\ref{eq:peeadiab}), while $p-p$ and $^7$Be neutrinos are in the vacuum averaged regime
(Equation~\ref{eq:peevacuum}).

\subsection{Experimentally Determined Solar Neutrino Parameters}
\label{subsec:parameters}

All of the results discussed in this section are taken from an
analysis given in Ref.~\cite{roadmap} of all currently available
solar neutrino and reactor anti-neutrino experimental data. In
this analysis,  all solar neutrino fluxes are treated as free
parameters subject only to the restriction that the fluxes satisfy
the luminosity constraint. The  evolution in the Sun and in the
Earth of the neutrino wavefunctions is solved for numerically. The
luminosity constraint imposes energy conservation provided that
the Sun shines by nuclear fusion reactions among light
elements~\cite{luminosity}. Where numerical allowed intervals of a
given parameter are reported, we marginalize over all other
variables including $\theta_{13}$ and $\Delta M^2$ atmospheric. At
all points in oscillation parameter space, we use the value of all
other variables that minimizes $\chi^2$ for that set of
parameters.

The best-fit values and the $1\sigma$ uncertainties for $\Delta
m^2$ and $\tan^2 \theta_{12}$ are :
\begin{eqnarray}
\Delta m^2 = (7.3^{+0.4}_{-0.6}) \times 10^{-5} \, {\rm eV^2} \label{eqn:m} \\
\tan^2 \theta_{12} = 0.41 \pm 0.05 \label{eqn:theta}
\end{eqnarray}

In principle, $\nu_e$ could oscillate into a state that is a
linear combination of active ($\nu_a$) and sterile ($\nu_s$)
neutrino states ($ \nu_e \to \cos\eta \, \nu_a \,+\, \sin\eta
\,\nu_s $). The 1$\sigma$ allowed range for the active-sterile
admixture is
\begin{equation}
 \sin^2\eta \leq 0.10 \, .
 \label{eq:sterile}
\end{equation}
The result given in Equation~(\ref{eq:sterile}) implies that less
than 6\% of the $^8$B flux is in the form of sterile neutrinos in
the energy range observed by the Sudbury Solar Neutrino
Observatory.

Comparing the measured neutrino fluxes with the theoretical predictions, we find for BP04 :
\begin{eqnarray}
\phi({\rm pp})_{\rm measured} &=&(1.02 \pm 0.02 \pm 0.01)\phi({\rm pp})_{\rm theory} \label{eqn:pp} \\
\phi({^8{\rm B}})_{\rm measured} &=&(0.88 \pm 0.04 \pm 0.23) \phi({^8{\rm B}})_{\rm theory} \label{eqn:8b} \\
\phi({^7{\rm Be}})_{\rm measured} & = & (0.91^{+0.24}_{-0.62} \pm
0.11)\phi({^7{\rm Be}})_{\rm theory} \label{eqn:7be}
\end{eqnarray}
In Equation~(\ref{eqn:pp}) and Equation~(\ref{eqn:7be}), the $1\sigma$
experimental uncertainties are given before the $1\sigma$
theoretical uncertainties.

The  measured and  theoretical values for the fluxes agree within
their combined $1\sigma$ uncertainties. The measurement error of
the $^8$B neutrino flux is smaller than the uncertainty in the
theoretical calculation, but the opposite is true for the p-p and
$^7$Be neutrino fluxes.

The CNO fluxes are  poorly constrained by the available solar neutrino data
(see Ref.~\cite{cnopaper}). BP04 predictions of the CNO-generated luminosity
of the Sun (normalized to the measured photon luminosity) ,
$L_{\rm CNO} = 1.6 \pm 0.6\,\%$ are well inside the range
allowed experimentally, $L_{\rm CNO} = 0.0_{-0.0}^{+2.8}\,\%$.

The results described above were obtained using the hypothesis
that the Sun shines by nuclear fusion reactions among light
elements. From neutrino measurements alone, one can measure the
solar energy generation rate and then compare this neutrino
luminosity with the photon luminosity being radiated from the
solar surface.  This comparison would test the fundamental idea
that nuclear fusion reactions are responsible for the energy
radiated by the Sun. Moreover, this same comparison would test a
basic inference from the standard solar model, namely, that the
Sun is in a quasi-steady state in which the energy currently
radiated from the solar surface is currently balanced by the
energy being produced by nuclear reactions in the solar interior.
We find for the ratio of the neutrino-inferred solar luminosity,
$L_\odot{\rm (neutrino-inferred)}$, to the accurately measured
photon luminosity, $L_\odot$, that
\begin{equation}
 \frac{L_\odot{\rm
  (neutrino-inferred)}}{L_\odot}~=~1.4^{+0.2}_{-0.3} .
 \label{eq:lnuoverlphoton}
\end{equation}
The neutrino-inferred solar luminosity is still very uncertain at present. This result reflects once
more the need of better determined low energy neutrino fluxes.

What do we expect from larger data samples in running experiments?
A global analysis using simulated three years of data for KamLAND shows that the uncertainty
of $\Delta m^2$ (Equation \ref{eqn:m}) will be reduced  by a factor of 2.5 \cite{roadmap}. SNO neutral
current measurements ($^3$He counters) will be able to reduce the uncertainty of $\tan^2 \theta_{12}$
by a 20\%. The neutrino fluxes summarized above are not affected, to the accuracy shown, by the
additional simulated KamLAND data and improved SNO neutral current measurement.

\section{Neutrinos as dark matter}
\label{sec:darkmatter}

Neutrinos are the first cosmological dark matter to be discovered.
Solar and atmospheric neutrino experiments show that neutrinos
have mass but these oscillations experiments only determine the
differences between masses, not the absolute values. If we make
the plausible but unproven assumption that the lowest neutrino
mass, $m_1$, is much less than the square root of $\Delta m^2_{\rm
solar}$, then we can conclude that the mass of cosmological
neutrino background is dominated by the mass of the heaviest
neutrino. This heaviest neutrino mass is then determined by
$\Delta m_{\rm atmospheric}^2$. With this assumption the
cosmological mass density in neutrinos is
only~\cite{mcdonald,3nuupdate,hitoshicarlos}

\begin{equation}
\Omega_\nu ~=~ (0.0009 \pm 0.0001)\, , ~~\, m_1 << \sqrt{(\Delta
m^2_{\rm solar})}. \label{eq:omegacosmological}
\end{equation}
Although the mass density given in
Eq.~(\ref{eq:omegacosmological}) is small, it is of the same order
of magnitude as the observed mass density in stars and gas.

The major uncertainty in determining by neutrino experiments the
value of $\Omega_\nu$ is the unknown value of the lowest neutrino
mass.  It is  possible that  neutrino masses are nearly degenerate
and cluster around the highest mass scale allowed by direct
beta-decay experiments.  If, for example, all neutrino masses are
close to 1 eV, then $\Omega_\nu({\rm 1~ev}) \sim 0.03$, which
would be cosmologically significant.

More sensitive neutrino beta-decay experiments and neutrinoless
double beta-decay experiments offer the best opportunities for
determining the mass of the lowest mass neutrino and hence
establishing the value of $\Omega_\nu$ from purely laboratory
measurements.

\section{What can be learned from new solar neutrino experiments?}
\label{sec:newexperiments}

 We begin our discussion of new
solar neutrino experiments by presenting in
Section~\ref{subsec:whylowenergy} the four primary reasons for
doing low energy solar neutrino experiments. Next we discuss in
Section~\ref{subsec:be7}, Section~\ref{subsec:pp}, and
Section~\ref{subsec:pep}, respectively, what can be learned from
future $^7$Be, $p-p$, and $pep$ solar neutrino experiments.
Finally, we describe in Section~\ref{subsec:protondecay} what can
be learned from parasitic solar neutrino experiments that are
carried out in connection with a next generation proton decay
experiment. The material in
Section~\ref{subsec:whylowenergy}-Section~\ref{subsec:pep} is
based upon Ref.~\cite{roadmap}.

\subsection{Why do low energy solar neutrino experiments?}
\label{subsec:whylowenergy}

There are four primary reasons for doing low energy solar neutrino
experiments that measure the energy of individual neutrino-induced
events.

First, new phenomena may be revealed at low energies ($< 3$ MeV)
that are not discernible at high energies ($> 5$ MeV). According
to the currently accepted LMA oscillation solution, the basic
oscillation mechanism switches  somewhere in the vicinity of 2 MeV
(see Equation~\ref{eq:critical} and Figure~\ref{fig:survival})
from the MSW matter-dominated oscillations that prevail at high
energies to the vacuum oscillations that dominate at low energies.
Does this transition from matter-induced to vacuum oscillations
does actually take place? If the transition does occur, is the
ratio ($\beta$, see Equation~\ref{eq:defbeta} and
Equation~\ref{eq:betaconvenient}) of the kinematic term in the
Hamiltonian (i.e., $\Delta m^2/2E$) to the matter-induced
term($\sqrt{2} G_{\rm F} n_{\rm e}$) the only parameter that
determines the physical processes that are observed in this energy
range?

Second, new solar neutrino experiments will provide accurate
measurements of the fluxes of the important $p-p$ and $^7$Be solar
neutrino fluxes, which together amount to more than 98\% of the
total flux of solar neutrinos predicted by the standard solar
model.  These measurements will test the solar model predictions
for the main energy-producing reactions, predictions that are more
precise than for the higher-energy neutrinos. Using only the
measurements of the solar neutrino fluxes, one can determine the
current rate at which energy is being produced in the solar
interior and can compare that energy generation rate with the
observed photon luminosity emitted from the solar surface. This
comparison will constitute a direct and accurate test of the
fundamental idea that the Sun shines by nuclear reactions among
light elements. Moreover, the neutrino flux measurements will test
directly a general result of the standard solar model, namely,
that the Sun is in a quasi-steady state in which the interior
energy generation rate equals the surface radiation rate.

Third, future solar neutrino experiments will make possible a
precise measurement of the vacuum mixing angle, $\theta_{12}$,  as
well as a slightly improved constraint on $\theta_{13}$. The
increased robustness in determining mixing angles will be  very
useful in connection with searches for CP violation. Uncertainties
in the CP-conserving neutrino parameters could compromise the
determination of the CP violating phase.

Fourth, there may be entirely new physical phenomena that show up
only at the low energies, the very long baseline,  and the great
sensitivity to matter effects provided by  solar neutrino
experiments. The reader will recall that solar neutrino research
was initiated to study the solar interior, not to search for
neutrino oscillations. Recently, two possibilities have been
discussed in which  new physics that is compatible with all
present data could show up at low energies in solar neutrino
experiments.  1). There could be a  sterile neutrino
 with very small mixing to active neutrinos and with a mass splitting smaller than the LMA
splitting~\cite{sterile}. Matter effects in the Sun would
resonantly enhance the mixing in vacuum producing at energies
around 1~MeV a much stronger deficit than pure LMA oscillations.
2). There could be small flavor-changing neutrino-matter
interactions~\cite{nsi}. These extra interactions would profoundly
modify the conversion probability at energies lower than around 6
MeV. Either mechanism would have strong particle physics
implications.

In this paper, we have assumed the correctness of all solar
neutrino and reactor experiments that have been performed so far
or which will be performed in the future. But, the history of
science teaches us that this is a dangerous assumption. Sometimes,
unrecognized systematic uncertainties can give misleading results.
To be sure that our conclusions are robust, the same quantities
must be measured in different ways.

\subsection{A $^7$Be experiment}
\label{subsec:be7}

The existing solar plus reactor experiments provide only loose constraints on the $^7$Be solar
neutrino flux, corresponding to approximately a $\pm 40$\% uncertainty at $1\sigma$.
We need an experiment to measure directly the  flux of $^7$Be solar neutrinos!

How accurate does the $^7$Be experiment have to be in order to provide important new information?
A measurement of the $\nu-e$ scattering rate accurate to $\pm 10$\% or better will reduce by a factor
of four the uncertainty in the measured $^7$Be neutrino flux. Moreover, the $10$\% $^7$Be flux
measurement will reduce the uncertainty in the crucial $p-p$ flux by a factor of about 2.5. That
improved determination of the $p-p$ flux by a $^7$Be measurement is due to the luminosity constraint.
A $^7$Be measurement accurate to $\pm 3$\% would provide another factor of two improvement in the
accuracy of the $^7$Be and $p-p$ solar neutrino fluxes.

All of these improvements are measured with respect to what we expect can be achieved with three years of
operation of the KamLAND experiment. Comparable information can be obtained from a CC (neutrino
absorption) experiment and from a neutrino-electron scattering experiment if both are performed to
the same accuracy.

Contrary to what some authors have stated, a $^7$Be solar neutrino experiment is not expected to provide
significantly more accurate values for the neutrino oscillation parameters than what we think will be
available after three years of operation of KamLAND.

\subsection{A p-p experiment}
\label{subsec:pp}

According to the standard solar model,about 91\% of the
total flux of the neutrinos from the Sun is in the form of the low energy ($<0.42$ MeV) $p-p$ neutrinos.
We cannot be sure that we have an essentially correct description of the solar interior until this
fundamental prediction is tested. Moreover, the $p-p$ neutrinos are in the range where vacuum
oscillations dominate over matter effects, so observing these low-energy neutrinos is an opportunity to
test in a crucial way also our understanding of the neutrino physics.

If we really know what we think we know, if the standard solar
model is correct to the stated accuracy ($\pm 1$\% for the total
$p-p$ neutrino flux), and if there is no new physics that shows up
below $0.4$ MeV, then a measurement of the $p-p$ flux to an
accuracy of better than $\pm 3$\% is necessary in order to
significantly improve our experimental knowledge of $\tan^2
\theta_{12}$. The main reason why such high accuracy is required
is that the existing experiments, if they are all correct to their
quoted accuracy, already determine the $p-p$ solar neutrino flux
to $\pm 2$\%.  (We assume that three years of KamLAND reactor data
will be available, as well as a $\pm 5$\% measurement of the
$^7$Be neutrino-electron scattering rate.)

As described above, an accurate measurement of the $p-p$ solar
neutrino flux will provide a direct test of the fundamental ideas
underlying the standard solar model. The $p-p$ measurement will
make possible the determination of the total solar luminosity from
just neutrino experiments alone. The neutrino luminosity can be
compared with the photon luminosity to check whether nuclear
fusion reactions among light elements is the only discernible
source of solar energy and whether the Sun is in an approximate
steady state in which the rate of interior energy generation
equals the rate at which energy is radiated through the solar
surface. The global combination of a $^7$Be experiment, plus a
$p-p$ experiment, plus the existing solar data, and three years of
KamLAND would make possible a precise determination of the solar
neutrino luminosity. A $p-p$ solar neutrino experiment accurate to
5\% would make possible a measurement of the solar neutrino
luminosity to 4\% and a 1\% $p-p$ experiment would determine the
solar luminosity to the accuracy implied below:
\begin{equation}
 \frac{L_\odot{\rm
  (neutrino-inferred)}}{L_\odot}~=~0.99 \pm 0.02 .
 \label{eq:lnuoverlphotonbepp}
\end{equation}

\subsection{A pep experiment}
\label{subsec:pep}

Assuming that the $pep$ neutrino flux (a $1.4$ MeV neutrino line)
is measured instead of the $p-p$ neutrino flux, we repeated the
global analyses of existing and future solar and KamLAND data. The
global analyses show that a measurement of the $\nu-e$ scattering
rate by $pep$ solar neutrinos would yield essentially equivalent
information about neutrino oscillation parameters and solar
neutrino fluxes as a measurement of the $\nu-e$ scattering rate by
$p-p$ solar neutrinos. The estimated best-estimates and
uncertainties in the parameters are almost identical for the
analyses we have carried out for $p-p$ and $pep$ neutrinos.

\subsection{Proton decay experiments and solar neutrino measurements}
\label{subsec:protondecay}

Large water Cherenkov detectors can make a unique and important
test of mater oscillations using $^8$B solar neutrinos. Only a
very large detector will have an event rate that is sufficiently
high to detect with statistical confidence the day-night effect
with solar neutrinos, an effect which is a characteristic signal
of matter-induced neutrino oscillations (the MSW effect).

Motivated by the UNO proposal~\cite{uno}, we suppose for
specificity that a future Cherenkov detector will have a fiducial
volume seven times that of Super-Kamiokande and that this detector
can measure neutrino-electron scattering above 6 MeV. We also
assume that the backgrounds and the photo-multiplier coverage
($\sim 40$\%) will be similar to the Super-Kamiokande experiment.

The best-fit LMA solution predicts a 2 \% day-night difference in
$\nu-e$ scattering event rates, which can be observed as a
$4\sigma$ effect in approximately ten years. A water Cherenkov
proton decay experiment would also provide a much more precise
measurement (much better than 1 \%) of the total event rate for
the scattering of $^8$B solar neutrinos by electrons.

A first detection of the very rare but high energy $hep$ neutrinos
should also be possible. We estimate that a measurement of the
$hep$ flux with the hypothesized proton decay detector should
achieve a $4\sigma$ or better accuracy over ten years.  This
result assumes that the BP04 predicted $hep$ flux is correct.

For these measurements of solar neutrinos to be successful, the
proton decay detector should be placed at a good depth with an
active shield. Special care should be taken to make sure that
radon contamination is low. Frequent calibrations should be made
to ensure that the detector sensitivity and the detector threshold
do not vary significantly, in an unknown way, from day to night.
The procedure for performing the day-night calibrations should be
included in the planning for the next generation proton decay
detector.

The study of solar neutrinos with large water Cherenkov detectors
is an ideal complement to the study of nucleon decay. The event
rate for nucleon decay cannot be predicted with confidence,
although the importance of just one or a few events is enormous.
The event rate for $^8$B solar neutrinos can be predicted with
great confidence and is enormous, about 31,100  events per year.

Somewhat paradoxically, the study of $^8$B solar neutrinos could
turn out to be the bread and butter project of next generation
water Cherenkov proton decay detectors.

 \ack JNB and CPG acknowledge support from NSF grant No.~PHY0070928.

\end{document}